\documentclass{article}
\usepackage{ijcai26}

\usepackage{times}
\usepackage{soul}
\usepackage{url}
\usepackage[hidelinks]{hyperref}
\usepackage[utf8]{inputenc}
\usepackage[small]{caption}
\usepackage{graphicx}
\usepackage{amsmath}
\usepackage{amsthm}
\usepackage{booktabs}
\usepackage{algorithm}
\usepackage{algorithmic}
\usepackage[switch]{lineno}

\usepackage{textcomp}
\usepackage{verbatim}
\usepackage{fancyhdr}
\usepackage{booktabs}
\usepackage{svg}
\usepackage{enumitem}
\usepackage{multirow}
\usepackage{mathrsfs}
\usepackage{amsmath}
\usepackage{makecell}
\usepackage{booktabs}
\usepackage{amssymb} 

\title{Proactive Guiding Strategy for Item-side Fairness in Interactive Recommendation}

\author{
Chongjun Xia$^{1,2}$
\and
Xiaoyu Shi$^1$\and
Hong Xie$^3$\\
Xianzhi Wang$^2$\and
Yun Lu$^1$\And
Mingsheng Shang$^1$
\affiliations
$^1$Chinese Academy of Sciences\\
$^2$University of Technology Sydney\\
$^3$University of Science and Technology of China\\
\emails
a260813501@gmail.com
}

\begin{document}

\maketitle

\begin{abstract}
Item-side fairness is crucial for ensuring the fair exposure of long-tail items in interactive recommender systems. Existing approaches promote the exposure of long-tail items by directly incorporating them into recommended results. This causes misalignment between user preferences and the recommended long-tail items, which hinders long-term user engagement and reduces the effectiveness of recommendations. We aim for a proactive fairness-guiding strategy, which actively guides user preferences toward long-tail items while preserving user satisfaction during the interactive recommendation process. To this end, we propose HRL4PFG, an interactive recommendation framework that leverages hierarchical reinforcement learning to guide user preferences toward long-tail items progressively. HRL4PFG operates through a macro-level process that generates fairness-guided targets based on multi-step feedback, and a micro-level process that fine-tunes recommendations in real time according to both these targets and evolving user preferences. Extensive experiments show that HRL4PFG improves cumulative interaction rewards and maximum user interaction length by a larger margin when compared with state-of-the-art methods in interactive recommendation environments.
\end{abstract}

\section{Introduction}
\label{sec:1}
Interactive recommender systems are an essential tool for modern online platforms to deliver personalized content, suggest relevant items, and drive user engagement in real time~\cite{he2016interactive}. These systems track and model evolving user preferences to improve user experience across diverse domains, from e-commerce to live-streaming platforms~\cite{quanavlrecom2023}.
Despite their effectiveness, recommender systems face fairness challenges due to popularity biases~\cite{abdollahpouri2019unfairness}, i.e.,
a small portion of popular items attract most of the attention, while many less popular (i.e., long-tail) items suffer from poor exposure, diminishing their visibility to users.

\begin{figure}[!t]
	{
		\centering
		\includegraphics[width=0.456\textwidth]{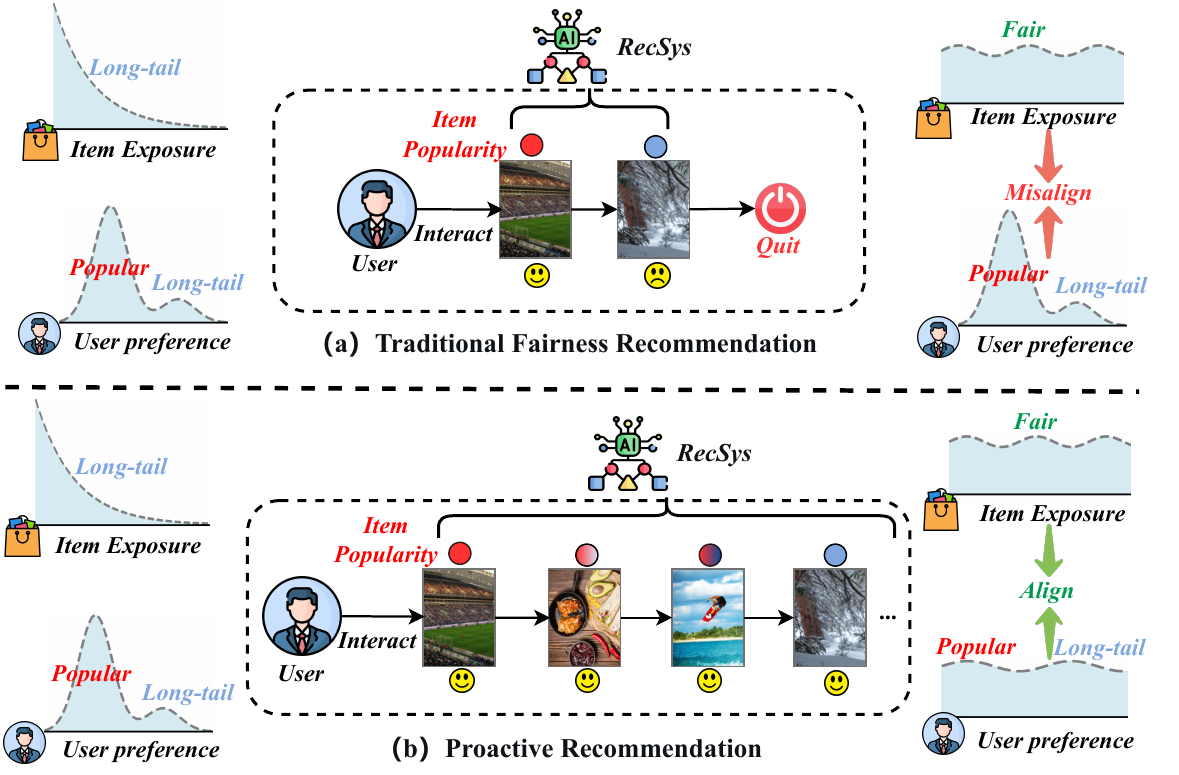}
		\caption{Illustration of two types of item-side fairness-aware recommendation.}
		\label{fig:motivation}
	}
\end{figure}

In this work, we focus on \emph{item-side fairness}, which we define as a more equitable allocation of exposure opportunities across items. This notion of fairness is particularly relevant in content platforms where creators, sellers, or information sources act as stakeholders whose utilities are directly tied to item exposure~\cite{wang2023survey}.
Within this perspective, prior work has primarily sought to mitigate the exposure disadvantage of long-tail items~\cite{ref22}.
They leverage causal inference techniques to mitigate popularity bias~\cite{ref16}, incorporate fairness-aware regularization terms during model training~\cite{ref4}, or employ re-ranking strategies to adjust recommendation outputs~\cite{zhu2021fairness}.
For dynamic environments, existing work often designs effective reward functions or incorporates fairness constraints into Reinforcement Learning (RL) to improve long-tail item exposure~\cite{shi2023relieving,shi2024maximum}.  
All these approaches promote the exposure of long-tail items by directly blending them into recommendations.   
Such an abrupt fairness-enforcement strategy disrupts the alignment between the recommended items and user preferences, resulting in reduced user engagement and satisfaction.
Figure~\ref{fig:motivation}(a) illustrates a traditional fairness-aware recommendation scenario, where the recommendation model prioritizes the exposure of long-tail items immediately after users interact with popular items. 
While the approach improves the distribution of long-tail item exposure, 
this abrupt shift in recommendations could appall users, discourage them from engaging, and ultimately, diminish user satisfaction.

Recent studies introduce the concept of \textit{Proactive Recommendation}, which suggests recommender systems should gradually guide the evolution of user preferences over time~\cite{tang2019target,wu2019proactive,zhu2023influential}. Inspired by this observation, we propose a proactive fairness-guiding strategy to enhance item-side fairness while preserving user satisfaction in interactive recommendation. Different from existing methods that force the exposure of long-tail items without aligning user preferences and recommended items, our strategy generates personalized and actionable recommendations iteratively throughout the interactive process. In this way, our approach progressively adjusts user preferences with the characteristics of long-tail items, thus achieving item-side fairness without impairing user engagement.
Figure \ref{fig:motivation}(b) illustrates an interactive recommender with our proactive fairness-guiding strategy, which gradually reduces the exposure of popular items based on the user’s current preferences, guiding users to appreciate long-tail items over time. This enables user preferences to transition gradually from a long-tail distribution to a more balanced distribution without compromising long-term satisfaction.

Implementing a proactive fairness-guiding strategy poses several challenges: 1) User preferences evolve during users' dynamic interactions with recommendation systems. Static or rule-based approaches fail to accommodate this fluidity, thus leading to negative user experiences. 2) Randomly selecting long-tail items for item-side fairness can result in recommending low-quality items, which undermines the overall effectiveness of the system. 3) Striking a balance between guiding user preference toward long-tail items and ensuring user satisfaction remains a complex and unresolved challenge in interactive recommendation research.

In this paper, we introduce Hierarchical Reinforcement Learning for Proactive Fairness-Guiding (HRL4PFG), a novel interactive recommendation framework that leverages reinforcement learning to address these challenges. HRL4PFG employs a hierarchical learning approach, dividing the guiding process into two phases: macro-learning and micro-learning. In the macro-learning phase, the high-level agent formulates fairness-guided targets based on multi-step feedback, considering both item-side fairness and user satisfaction. These targets guide the micro-learning phase. In the micro-learning phase, the low-level agent translates these high-level targets into executable actions tailored to individual users. The low-level agent balances the trade-off between achieving defined targets and maintaining user satisfaction, gradually steering user preference toward the desired target. Through this progressive and adaptive guiding strategy, HRL4PFG enhances item-side fairness in recommender systems while minimizing disruptions to user satisfaction. In a nutshell, this work makes the following contributions:
\begin{itemize}
   \item We emphasize the necessity of using a proactive fairness-guiding strategy to gradually influence user preference for long-tail items and utilize it for enhancing item-side fairness in the interactive recommendation.
   \item We propose HRL4PFG, a hierarchical reinforcement learning framework to improve item-side fairness in interactive recommender systems. HRL4PFG employs a hierarchical learning approach, dividing the guiding process into two tasks: the high-level agent generates fairness-guided targets, and the low-level agent selects recommended items based on these targets.
   \item We conduct extensive interactive experiments to validate the effectiveness and superiority of HRL4PFG, in terms of promoting maximum interaction length and cumulative interaction reward. 
\end{itemize}

\section{Related Work}
\label{sec:2}
\subsubsection{Fairness in Recommendation.}
Fairness has drawn increasing attention in recommendation research, aiming to reduce popularity biases and inequalities~\cite{ref19}.
Fairness is typically considered at the group level or the individual level. Group fairness ensures equitable treatment across demographic groups, and individual fairness ensures similar users and items are treated equally~\cite{abdollahpouri2019unfairness,singh2019policy}.

Several methods address fairness in recommender systems~\cite{ref16,ref22,zhu2021fairness}.
For example, Abdollahpouri et al.~\cite{ref4} propose a regularization term to reduce popularity bias;
Zhu et al.~\cite{zhu2018fairness} introduce a fairness-enhancing framework with a latent factor matrix and fairness regularization. Both the above methods have focused on the fairness issue in static, one-shot recommendation settings and optimizing fairness at a single point in time.
As such, they overlook the dynamicity of user-system interactions and are unsuitable for achieving item-side fairness in interactive recommendation processes.

\subsubsection{Proactive Recommendation.}
Proactive recommendation departs from traditional recommendation paradigms by shifting from passively adapting to user preference towards actively guiding user preference in a desired direction~\cite{bi2024proactive}. This proactive approach offers several benefits, such as mitigating the filter bubble problem and enabling effective advertisement placement. Research on proactive recommendation focuses on two primary directions: 1) modeling preference shifts within recommendation systems, and 2) guiding user preferences proactively.

Models of preference shifts study how user interests evolve over interactions with recommender systems~\cite{curmei2022towards,dean2022preference}. In contrast, preference-guidance methods aim to smoothly steer users toward target items/topics while preserving satisfaction~\cite{tang2019target,wu2019proactive,zhu2023influential}.
However, most prior work does not clearly specify how guiding targets are chosen. We address item-side fairness in interactive recommendation by generating fairness-oriented objectives and applying a smoothing guidance strategy accordingly.

\subsubsection{Reinforcement learning for Recommendation.}
Reinforcement learning (RL) has gained increasing attention in recent years for addressing the dynamics of interactive recommendation scenarios~\cite{bharadhwaj2019meta}. Traditional recommendation algorithms, which typically optimize static, one-shot objectives such as maximizing click-through rates, fail to account for the evolving preferences and behaviors of users~\cite{zhang2019deep,zhang2025sarrec}. In contrast, RL models the recommendation process as a sequence of decisions within a dynamic environment, enabling continuous adaptation to changes in user preference~\cite{zheng2018drn}. Recently, hierarchical reinforcement learning (HRL) has emerged as a powerful method across various domains, including recommender systems~\cite{ref7,ref9}.  In this work, we adopt HRL to enhance item-side fairness in interactive recommender systems.

Several studies explored the use of reinforcement learning to address item-side fairness in interactive recommender systems by modifying the reward function or adding constraints~\cite{shi2023relieving,shi2024maximum}. However, directly recommending long-tail items to users who prefer popular content can severely disrupt the alignment between recommended items and user preference, resulting in reduced engagement and satisfaction. Since user preference can be gradually influenced by recommender systems~\cite{zhu2023influential}, our method seeks to progressively guide users toward long-tail items. This gradual approach aims to increase the exposure of long-tail items while maintaining user satisfaction.

\section{Methodology}
\label{sec:3}


\subsection{Task Formulation}
Let \(u\in\mathcal{U}\) and \(i\in\mathcal{I}\) denote a user and an item.
In interactive recommendation, the system interacts with users over trajectories.
At step \(t\), it observes a user state \(s_t^u\) summarized from the interaction history and recommends an item \(i_t^u\), receiving feedback \(r_t^u\).
User preferences evolve with interactions, leading to non-stationary states over time.

\textbf{Item-side fairness.}
We focus on item-side fairness, which aims to allocate exposure opportunities more equitably across items at the system level~\cite{wang2023survey}. In particular, it emphasizes mitigating the exposure disadvantage of long-tail items.
Let \(\mathcal{I}_{\mathrm{tail}}\subset\mathcal{I}\) be the long-tail set.
Given a collection of interaction trajectories over a user set \(\mathcal{U}\), define the total exposure of item \(i\) as:
\begin{equation*}
\mathrm{Exp}(i) \;=\; \sum_{u\in\mathcal{U}}\sum_{t=1}^{T_u}\mathbb{I}\!\left[i_t^u=i\right].
\end{equation*}
Accordingly, the long-tail exposure ratio is calculated as:
\begin{equation*}
\rho \;=\; \frac{\sum_{i\in\mathcal{I}_{\mathrm{tail}}}\mathrm{Exp}(i)}
{\sum_{i\in\mathcal{I}}\mathrm{Exp}(i)},
\end{equation*}
which measures the percentage of total exposure that is allocated to long-tail items across all interaction steps.

\textbf{Proactive fairness-guided recommendation.}
To improve \(\rho\) without degrading user satisfaction, the system adopts a proactive guiding strategy.
Specifically, it generates a fairness-guided target \( j \) (i.e., a virtual long-tail item) and steers user preference toward this target over $M$ interactive recommendation steps. 
The goal is to learn a policy \(\pi\) that maximizes long-term user satisfaction while increasing system-level long-tail exposure.

\subsection{Framework Overview}
\begin{figure*}[t]
    {
      \centering
      \includegraphics[width=0.906\textwidth]{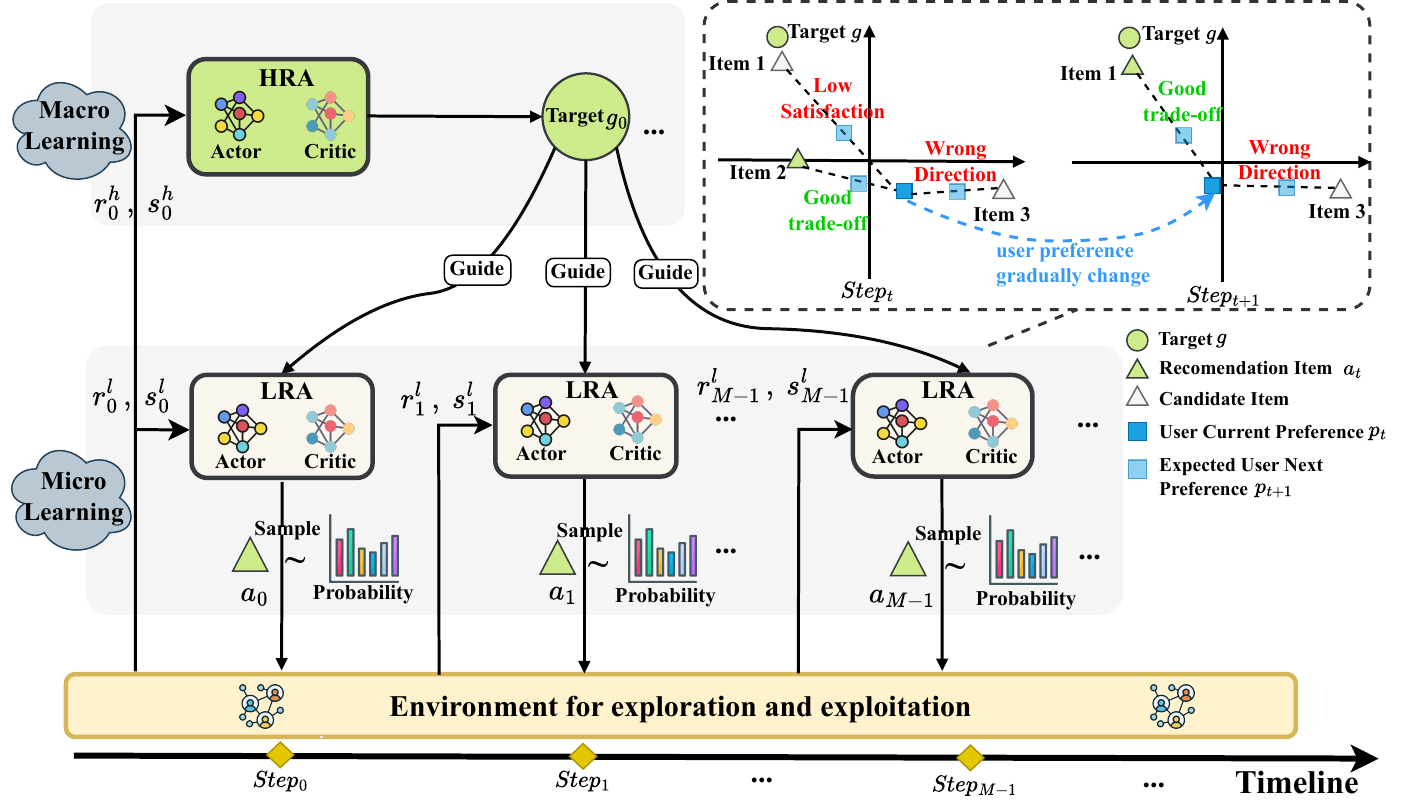}
      \caption{The overall architecture of HRL4PFG.}
      \label{fig:overall_framework}
    }
\end{figure*}

Based on the above settings, we propose HRL4PFG, a novel interactive recommendation framework that leverages hierarchical reinforcement learning to improve item-side fairness. HRL4PFG splits the guiding process into two phases (as shown in Figure \ref{fig:overall_framework}): 

\textbf{1) Macro-learning:} the high-level agent (HRA) consists of a high-level actor and a critic. The actor is responsible for generating fairness-guided targets, while the critic evaluates these targets based on multi-step user feedback and item-side fairness of the recommendations. Based on this evaluation, the critic iteratively refines the actor’s decision-making process.

\textbf{2) Micro-learning:} the low-level agent (LRA) comprises a low-level actor and a critic. The actor generates item recommendations by considering both the user’s current preference and fairness-guided targets, aiming to progressively align user preference with the fairness targets. The critic evaluates the effectiveness of these recommendations in guiding user preference while assessing their alignment with the user’s current preference. Based on these evaluations, the critic optimizes the actor's decision-making process.

In the following subsections, further details of each component in the HRL4PFG are provided.

\subsection{Macro Learning}
The macro-learning process involves the generation of a fairness-guided target by the high-level agent (HRA) to enhance fair item exposure. This fairness-guided target serves as a reference for directing actions of the low-level agent over the next $M$ steps.
\subsubsection{High-level State Tracker.}
The high-level state \( s_t^h \) contains information about the most recent \( N \) user interactions \( \{h_{t-N+1}, \dots, h_t\} \), reflecting the user's preference. To model the dependencies among these interactions, we incorporate an attention mechanism. Specifically, the user interactions are first encoded into vector representations \( \{e_{t-N+1}, \dots, e_t\} \) through an embedding layer. Then, all historical embedding vectors are mapped into three matrices: query matrix \( Q^h \), key matrix \( K^h \), and value matrix \( V^h \). Finally, the high-level state representation can be computed as follows:
\begin{equation} 
s_t^h = Attention(Q^h, K^h, V^h) = softmax\left( \frac{Q^h {(K^h)}^T}{\sqrt{d^h}} \right) V^h,
\label{eq:high_level_tracker}
\end{equation}
where $d^h$ is a scaling factor.

\subsubsection{High-level Agent Model.}
The high-level actor is designed to generate a dynamic, fairness-guided target that steers user preference toward a more equitable exposure of items. To facilitate comprehensive exploration and prevent premature convergence, we employ a Gaussian distribution. Specifically, the high-level state \( s_t^h \) is input into a multilayer perceptron to generate the distribution's mean \( u_t^h \) and variance \( \sigma_t^2 \).

Subsequently, the fairness-guided target \( g_t \) is sampled from the Gaussian distribution parameterized by the mean \( u_t^h \) and variance \( \sigma_t^2 \):
\begin{equation}
    g_t \sim \mathcal{N}(u_t^h, \sigma_t^2).
\end{equation}

The high-level agent is implemented using the Actor-Critic algorithm\cite{konda1999actor}. The advantage of the Actor-Critic algorithm lies in its stable learning process, reduced variance, and efficient policy optimization by combining policy gradient and value-based methods. The critic receives \(s^h_t\) and estimates the state return \(V_{\varphi^h} (s_{t}^h)\). The high-level critic networks are trained by minimizing the following loss function:
\begin{equation}
\label{eq:loss_ac1}
	\mathcal{L}_{v}^h= \mathbb{E}_t [(V_{\varphi^h}(s_t^h) - R^h_t)^2], \quad R^h_t  =\gamma V_{\varphi_{\_}^h}(s_{t+1}^h) + r_t^h,
\end{equation}
where $\varphi^h$ and $\varphi_{\_}^h$ are parameters of the online and target network in the high-level critic. The discount factor $\gamma \in [0, 1]$ is used to balance the importance of immediate rewards and future rewards. The main network of the actor is trained by minimizing the loss function:
\begin{equation}
\label{eq:loss_ac2}
\mathcal{L}_{a}^h = -\mathbb{E} \left[ \log \pi_{\theta^l}(g_t | s_t^h) \cdot A^h(s_t^h, g_t) \right],
\end{equation}
where $A^h$ is the advantage function and can be calculated as follow:
\begin{equation}
\begin{aligned}
A^h(s_t^h, g_t)= \gamma V_{\varphi_{}^h} (s_{t+1}^h) + r_{t}^h-V_{\varphi^h_{}} (s_{t}^h),
\end{aligned}
\end{equation}

\subsubsection{High-level Reward Mechanism.} 
The fairness-guided target \( g_t \) is employed to guide the LRA over next \( M \) steps. However, not every generated \( g_t \) is meaningful, as it may deviate significantly from the true item space.
To address this, we introduce the following constraint in the reward function:
\begin{equation}
\label{eq:condition} 
\|g_t - v_{center}\|_2 \leq \max_{i \in \mathcal{I}} \left( \|v_i - v_{center}\|_2 \right),
\end{equation}
where \( \mathcal{I} \) denotes the set of items, \( v_i \) is the encoding vector of item \( i \), and \( v_{\text{center}} \) represents the center point of all item encodings. The function \( \|\cdot\|_2 \) denotes the \( L2 \) norm, i.e., the Euclidean distance. Additionally, the reward function for the high-level agent should consider both recommendation accuracy and item-side fairness over the next $M$ steps. The formula for the reward function is given as follows:
\begin{equation} 
\label{eq:h_reward} 
r_t^{h} = \begin{cases} 
\sum_{j=t}^{t+M} \left( r_j^a + \lambda_f \cdot r_j^f \right), & \text{if Equation} (\ref{eq:condition}) \text{ is ture} \\ 
0, & \text{otherwise} 
\end{cases}
\end{equation}
where \( r_j^a \) represents the accuracy reward for the item recommended at time \( j \), which can be derived from user feedback such as ratings or click counts. \( r_j^f \) denotes item-side fairness reward and is defined as:
\begin{equation}
r_j^f = -\log(pop_j), \quad pop_j = \frac{N_j}{|U|},
\end{equation}
where \( N_j \) is the number of users who provided positive feedback for the item recommended at time \( j \), and \( |U| \) is the total number of users. The parameter \( \lambda_f \) is a weighting factor that balances recommendation accuracy and fairness.

\subsection{Micro Learning}
The low-level agent (LRA) selects recommended items based on the fairness-guided target generated by the HRA and the user's current preference. As illustrated in Figure~\ref{fig:overall_framework}, during each recommendation step, the LRA aims to steer user preference toward the fairness-guided target $g_t$ while ensuring sustained user satisfaction. Therefore, at $Step_{t}$, Item 2 is the optimal choice, while at $Step_{t+1}$, Item 1 becomes the optimal recommendation.

\subsubsection{Low-level State Tracker.}
The low-level state $s_t^l$ not only contains information about the user's recent interactions, which reflects the user's preference, but also includes a fairness-guided target that directs the system towards promoting fair item exposure. Similar to HRA, it first maps the most recent \(N\) user interactions \(\{h_{t-N+1},\dots, h_t\}\) through an embedding layer and then projects the embedding vectors into query matrix \(Q^l\), key matrix \(K^l\), and value matrix \(V^l\). Finally, an attention layer is used to extract user preference:
\begin{equation} 
p_t = Attention(Q^l, K^l, V^l) = softmax\left( \frac{Q^l {(K^l)}^T}{\sqrt{d^l}} \right) V^l,
\label{eq:high_level_tracker}
\end{equation}
where \( d^l \) is a scaling factor. The preference embedding \( p_t \) is concatenated with the fairness-guided target $g_t$ to obtain the low-level state: $s_t^l = p_t \oplus g_t$.

\subsubsection{Low-level Agent Model.} 
The low-level actor is responsible for generating recommended items that align with the user's current preference and guide user preference toward the fairness-guided target \( g_t \). However, the vast item space presents a challenge, as many items deviate from the intended direction dictated by the fairness-guided target. To enable efficient learning for LRA, we apply a \textbf{filtering mechanism} to the candidate items. First, we compute the $L2$ distance between the encoding of each item and the fairness-guided target \( g_t \), and then select the top \( L \) items that are closest to \( g_t \) as candidate items. The probability of each item being recommended is computed as follows:
\begin{equation}
\label{eq:l_action}
    P(\mathcal{I} \mid s_t^l) = \phi\left( MLP(s_t^l) \cdot Mask \right),
\end{equation}
where \( MLP \) represents a multilayer perceptron, and \( \mathcal{I} \) is the entire item space. The \( Mask \) is a masking vector where the positions of the selected candidate items are set to 1, and all other positions are set to 0. \( \phi(\cdot) \) is a normalization operation. 

After deriving the distribution $P(\mathcal{I}\mid s_t^l)$, items are sampled to form a list that satisfies user preference while promoting item-side fairness. The low-level agent adopts the Actor-Critic method, with loss functions similar to Equations (\ref{eq:loss_ac1}) and (\ref{eq:loss_ac2}).

\subsubsection{Low-level Reward Mechanism.}
An invalid fairness-guided target \( g_t \), which deviates significantly from the true item space, undermines the effectiveness of the low-level agent’s learning process. Consequently, we also integrate the constraint given in Equation~\ref{eq:condition} into the low-level reward function. Moreover, the recommendations of the low-level agent should both accurately align with the user's current preference and guide the evolution of user preference toward the desired target \( g_t \). To fulfill these objectives, the low-level reward function is designed as follows:
\begin{equation} 
\label{eq:l_reward} 
r_t^{l} = \begin{cases} 
r_t^{a} + \lambda_g \cdot r_t^g, & \text{if Equation} (\ref{eq:condition}) \text{ is true} \\ 
0, & \text{otherwise} 
\end{cases}
\end{equation}
\begin{equation}
  r_t^g = Distance(p_t, g_t) - Distance(p_{t+1}, g_t),
\end{equation}
where \( r_t^{a} \) represents the accuracy reward, which can be derived from user feedback such as ratings or click counts, while \( \lambda_g \) is a hyperparameter that controls the trade-off between user satisfaction and the guidance of user preference. The guiding reward component \( r_t^g \) is introduced to steer user preference in the direction specified by the fairness-guided target \( g_t \). The function \( Distance (\cdot) \) uses the \( L2 \) norm.

\section{Experiments}
\label{sec:5}
In this section, we conduct a series of experiments to address the following research questions related to HRL4PFG:
\begin{itemize}
  \item \textbf{RQ1}: How does HRL4PFG compare to other state-of-the-art methods in interactive recommendation setting?
  \item \textbf{RQ2}: How does HRL4PFG perform across various environments under different fairness constraints?
  \item \textbf{RQ3}: How effective are the different modules in HRL4PFG at improving recommendation performance?
  \item \textbf{RQ4}: How does HRL4PFG perform with different hyper-parameter settings?
\end{itemize}

\subsection{Experiments Setup}
In this section, we introduce the experimental settings, including details on the environment simulator, evaluation metrics, baseline methods, and parameters setting.

\subsubsection{Environment Simulator.}  
Due to the high cost of online experiments, we developed an interactive simulation environment based on the EasyRL4Rec library~\cite{yu2024easyrl4rec} to study methods for improving item-side fairness and maximizing users' long-term satisfaction in the interactive recommendation settings. We created two simulation environments using the recently proposed KuaiRec~\cite{gao2022kuairec} and KuaiRand~\cite{gao2022kuairand} datasets, which provide high-quality user interaction logs. We split the datasets as shown in Table \ref{tab:dataset_statistics} to construct separate training and testing environments. The testing environment remains unseen during the training of the recommendation models. 

\begin{table}[b]
  \centering
  \begin{tabular}{lcccc}
  \toprule
  \textbf{Dataset} & \textbf{Usage} & \textbf{Users} & \textbf{Items} & \textbf{Interactions} \\ 
  \midrule
  \multirow{2}{*}{KuaiRec} & Train & 7,176 & 10,728 & 12530.8k  \\ 
                           & Test  & 1,411 & 3,327  & 4676.5k   \\ 
  \midrule
  \multirow{2}{*}{KuaiRand} & Train & 26,285 & 7,551  & 1436.6k  \\ 
                            & Test  & 27,285 & 7,583  & 1186.1k  \\ 
  \bottomrule
  \end{tabular}
  \caption{Dataset statistics}
  \label{tab:dataset_statistics}
\end{table}

\begin{table*}[h]
  \centering
  \tabcolsep=0.16cm
  \renewcommand{\arraystretch}{1.0}
  \scriptsize
  \begin{tabular}{c|c|cccc|cccc}  
  \Xhline{1.8\arrayrulewidth}  
  \selectfont
  \multirow{2}{*}{Envs} & \multirow{2}{*}{Methods} & \multicolumn{4}{c|}{Max Len 30} & \multicolumn{4}{c}{Max Len 50} \\
  \cline{3-10}  
                        & & $\text{R}_\text{cum}$ $\uparrow$& $\text{R}_\text{single}$$\uparrow$ & Len $\uparrow$
                        & Gini Index (\%)$\downarrow$
                        & $\text{R}_\text{cum}$$\uparrow$ & $\text{R}_\text{single}$$\uparrow$ & Len $\uparrow$& Gini Index (\%)$\downarrow$
                        \\
                        
  \hline
  \multirow{8}{*}{KuaiRec} 
  &SQN  &13.46 $\pm$ 1.96  &\underline{1.00 $\pm$ 0.03}  &13.61 $\pm$ 2.35  &99.40 $\pm$ 0.10  &13.23 $\pm$ 4.32  &\underline{0.99 $\pm$ 0.04}  &13.65 $\pm$ 5.13  &99.30 $\pm$ 0.10\\
  &PG  &14.06 $\pm$ 1.23  &0.79 $\pm$ 0.08  &18.32 $\pm$ 2.95  &\underline{99.10 $\pm$ 0.20}  &17.76 $\pm$ 1.60  &0.71 $\pm$ 0.10  &26.02 $\pm$ 5.19  &98.70 $\pm$ 0.20\\
  &DDPG  &8.23 $\pm$ 3.99  &0.80 $\pm$ 0.22  &13.03 $\pm$ 8.59  &99.50 $\pm$ 0.20  &12.27 $\pm$ 4.41  &0.42 $\pm$ 0.17  &32.92 $\pm$ 10.29  &99.40 $\pm$ 0.30\\
  &TD3  &8.27 $\pm$ 5.68  &0.44 $\pm$ 0.28  &23.13 $\pm$ 8.29  &99.60 $\pm$ 0.30  &9.34 $\pm$ 6.69  &0.50 $\pm$ 0.34  &31.39 $\pm$ 19.43  &99.30 $\pm$ 0.60\\
  &C51  &13.90 $\pm$ 3.38  &0.77 $\pm$ 0.09  &18.84 $\pm$ 6.03  &99.00 $\pm$ 0.20  &19.12 $\pm$ 5.50  &0.76 $\pm$ 0.06  &25.61 $\pm$ 8.47  &\underline{98.40 $\pm$ 0.30}\\
  &SAC4IR  &4.86 $\pm$ 1.57  &0.25 $\pm$ 0.12  &23.39 $\pm$ 4.13  &99.70 $\pm$ 0.20  &9.60 $\pm$ 3.04  &0.26 $\pm$ 0.08  &\underline{38.21 $\pm$ 4.19}  &99.60 $\pm$ 0.30\\
  &DNAIR  &\underline{17.40 $\pm$ 3.09}  &0.84 $\pm$ 0.05  &\underline{21.04 $\pm$ 4.42}  &99.20 $\pm$ 0.20  &\underline{21.85 $\pm$ 5.58}  &0.82 $\pm$ 0.05  &{26.80 $\pm$ 6.96}  &98.90 $\pm$ 0.30\\
  &HRL4PFG  &\textbf{29.99 $\pm$ 2.98}  &\textbf{1.07 $\pm$ 0.04}  &\textbf{27.82 $\pm$ 2.34}  &\textbf{98.20 $\pm$ 1.30}  &\textbf{43.37 $\pm$ 4.00}  &\textbf{1.08 $\pm$ 0.03}  &\textbf{40.25 $\pm$ 3.63}  &\textbf{97.20 $\pm$ 1.10}\\
  \Xhline{1.8\arrayrulewidth} 
  \multirow{8}{*}{\quad KuaiRand\quad}
  &SQN  &1.74 $\pm$ 0.08  &\textbf{0.42 $\pm$ 0.02}  &4.11 $\pm$ 0.07  &99.90 $\pm$ 0.00  &1.76 $\pm$ 0.08  &\textbf{0.43 $\pm$ 0.02}  &4.07 $\pm$ 0.05  &99.90 $\pm$ 0.00\\
  &PG  &3.32 $\pm$ 0.51  &0.33 $\pm$ 0.02  &10.24 $\pm$ 1.63  &99.60 $\pm$ 0.10  &6.46 $\pm$ 0.70  &0.28 $\pm$ 0.01  &22.70 $\pm$ 2.35  &98.40 $\pm$ 0.70\\
  &DDPG  &5.41 $\pm$ 1.76  &0.26 $\pm$ 0.03  &21.23 $\pm$ 7.26  &99.70 $\pm$ 0.10  &6.96 $\pm$ 3.02  &0.25 $\pm$ 0.04  &29.20 $\pm$ 13.81  &99.60 $\pm$ 0.20\\
  &TD3  &4.35 $\pm$ 1.52  &0.26 $\pm$ 0.04  &18.50 $\pm$ 7.22  &99.60 $\pm$ 0.10  &6.09 $\pm$ 2.97  &0.27 $\pm$ 0.05  &25.61 $\pm$ 14.12  &99.60 $\pm$ 0.30\\
  &C51  &2.68 $\pm$ 0.62  &0.32 $\pm$ 0.03  &8.79 $\pm$ 2.77  &98.70 $\pm$ 0.20  &2.88 $\pm$ 0.99  &0.34 $\pm$ 0.02  &8.98 $\pm$ 4.08  &98.30 $\pm$ 0.30\\
  &SAC4IR  &\underline{6.21 $\pm$ 2.17}  &0.28 $\pm$ 0.03  &\underline{23.56 $\pm$ 8.12}  &99.70 $\pm$ 0.10  &5.85 $\pm$ 3.34  &0.28 $\pm$ 0.05  &24.00 $\pm$ 16.43  &99.70 $\pm$ 0.20\\
  &DNAIR  &5.97 $\pm$ 0.87  &0.26 $\pm$ 0.02  &22.78 $\pm$ 3.16  &\underline{98.10 $\pm$ 1.00}  &\underline{8.00 $\pm$ 1.36}  &0.27 $\pm$ 0.02  &\underline{30.54 $\pm$ 5.88}  &\underline{98.20 $\pm$ 0.80}\\
  &HRL4PFG  &\textbf{10.03 $\pm$ 1.85}  &\underline{0.36 $\pm$ 0.04}  &\textbf{27.90 $\pm$ 2.25}  &\textbf{95.50 $\pm$ 5.30}  &\textbf{15.41 $\pm$ 2.78}  &\underline{0.35 $\pm$ 0.03}  &\textbf{43.40 $\pm$ 4.83}  &\textbf{94.70 $\pm$ 6.60}\\
  \Xhline{1.8\arrayrulewidth} 
  \end{tabular}
  \caption{Average results of all methods in different environments.}
  \label{tab:overall_performance}
\end{table*}

The maximum interaction length (Max Len) is introduced to model realistic session-level recommendation scenarios, where users interact with a system for a bounded period of time. We consider Max Len = 30 and 50 to simulate medium- and long-horizon interactions. This means that a user can generate up to 30 or 50 interactions in each session. Similar to~\cite{shi2024maximum,yu2024easyrl4rec}, we introduce a popularity-aware exit mechanism to enable the simulation environment to assess the item-side fairness of the model's recommendations: if the recommendation model consecutively recommends $W$ items that rank in the top 20\% by popularity, the interaction terminates. A smaller value of $W$ implies a higher requirement for fairness in the model's recommendations.

\subsubsection{Evaluation Metrics.}
We evaluate model performance in the interactive recommendation setting along two dimensions, user satisfaction and item-side fairness.

\textbf{User Satisfaction Metrics.} Following~\cite{yu2024easyrl4rec}, we adopt \textbf{single-round reward} ($\text{R}_\text{single}$) and 
\textbf{cumulative reward} ($\text{R}_\text{cum}$) to measure immediate and long-term user satisfaction, respectively. $\text{R}_\text{single}$ reflects the quality of recommendations at each interaction step, while $\text{R}_\text{cum}$ captures the model’s ability to sustain high-quality interactions over an entire session. We further report \textbf{interaction length} (Len), which measures the number of interactions a user completes before exiting the system. 

\textbf{Item-side Fairness Metrics}. To explicitly quantify exposure fairness among items, we employ the \textbf{Gini Index}\cite{ge2021towards}, 
which measures inequality in item exposure distribution across the entire recommendation horizon. A lower Gini Index reflects improved item-side fairness.

\subsubsection{Baselines.}
To fairly and comprehensively evaluate the effectiveness of HRL4PFG, we compare it with both RL-based methods and fairness-aware RL approaches. The baselines include: SQN~\cite{xin2020self}, a dual-head model combining cross-entropy loss and RL for recommendation. PG~\cite{williams1992simple}, a policy gradient method that directly optimizes the policy by estimating reward gradients. DDPG~\cite{lillicrap2015continuous}, a deterministic policy gradient algorithm using actor-critic networks for optimization. TD3~\cite{fujimoto2018addressing}, an RL algorithm that reduces overestimation bias by using two critic networks and the minimum Q-value for stable learning. C51~\cite{bellemare2017distributional}, a distributional RL method that models the full reward distribution rather than expected values. SAC4IR~\cite{shi2024maximum}, a fairness-aware RL approach leveraging maximum entropy constraints to enhance long-term fairness. DNAIR~\cite{shi2023relieving}, an RL-based method integrating item popularity and quality to mitigate popularity bias.

SQN is trained using offline logs, while other algorithms employ a pre-trained user model (DeepFM~\cite{guo2017deepfm}) to guide the online agent's learning. All methods use a quit parameter $W = 3$ across environments. Key parameters are tuned for all methods. Detailed parameter settings and the code are provided in the Appendix.

\subsection{Overall Performance Comparison (RQ1)}
We conducted 50 training epochs for each method in different interaction environments. After learning in each epoch, we evaluate all methods with 100 episodes (i.e., interaction trajectories). The results are shown in Table~\ref{tab:overall_performance}, 

\paragraph{Item-side Fairness Analysis.} HRL4PFG consistently achieves the lowest Gini Index across both KuaiRec and KuaiRand under different interaction horizons. 
This indicates that HRL4PFG effectively reduces exposure inequality between popular and long-tail items.
Compared to the fairness-aware baselines, HRL4PFG further lowers the Gini Index, 
demonstrating the advantage of proactive, preference-aligned fairness guidance over reactive regularization-based approaches.

\paragraph{User Satisfaction under Fairness Constraints.}  Importantly, the improvement in exposure fairness does not come at the cost of user satisfaction. 
On the contrary, HRL4PFG simultaneously achieves the highest cumulative reward and the longest interaction length across all settings. 
This is because HRL4PFG adopts a progressive guidance mechanism that smoothly shifts user preferences toward long-tail items, thereby achieving effective alignment between item-side fairness objectives and user preferences.
Fairness-aware baselines like SAC4IR and DNAIR partially alleviate exposure imbalance, yet remain inferior to HRL4PFG in sustaining long-term engagement and cumulative reward.

\subsection{Results on Different Environments (RQ2)}
\begin{figure}[h]
  {
    \centering
    \includegraphics[width=0.446 \textwidth]{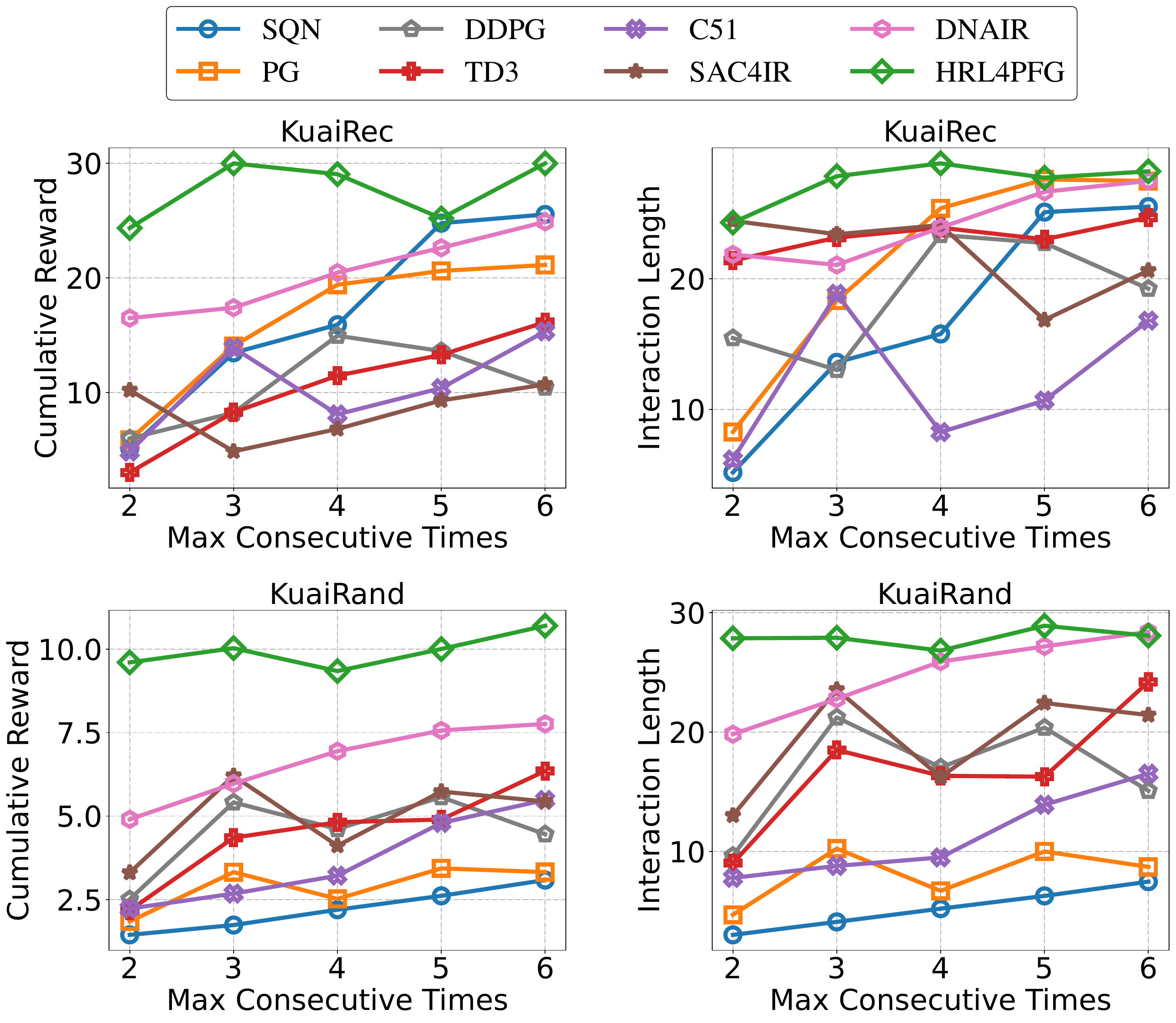}
    \caption{Results under different exit conditions.}
    \label{fig:robost}
  }
\end{figure}

To validate the robustness of HRL4PFG under different environmental settings, we varied the exit mechanism parameter $W$ to values of 2, 3, 4, 5, and 6, while keeping the maximum interaction length fixed at 30. A larger $W$ indicates a lower requirement for item-side fairness in the system. The experimental results are presented in Figure \ref{fig:robost}. Under different item-side fairness requirements, HRL4PFG consistently exhibits optimal performance, which demonstrates its robustness across diverse environments.

\subsection{Ablation Study (RQ3)}
To evaluate the effectiveness of each component in our proposed framework, we conduct an ablation study by defining the following model variants: 1) \textbf{HRL4PFG-wo-hie}: this variant removes the hierarchical structure by eliminating the high-level agent, leaving only the low-level agent for recommendation; 2) \textbf{HRL4PFG-wo-tc}: this variant removes the constraint on the fairness-guided target generated by the high-level agent, as described in Equation \ref{eq:condition}; 3) \textbf{HRL4PFG-wo-fm}: this variant removes the filtering mechanism $Mask$, as described in Equation \ref{eq:l_action}; 4) \textbf{HRL4PFG}: the full version of our proposed framework.

\begin{figure}[h]
  {
    \centering
    \includegraphics[width=0.446\textwidth]{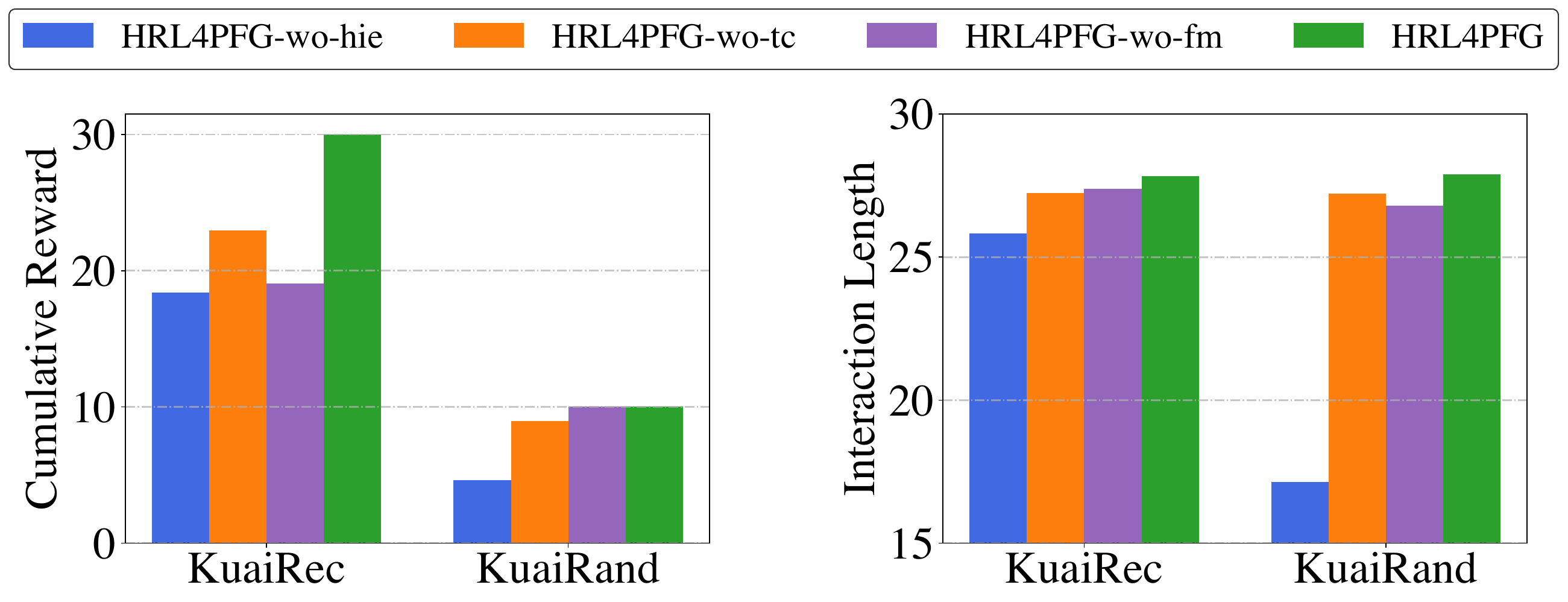}
    \caption{Ablation study results of HRL4PFG.}
    \label{fig:ablation}
  }
\end{figure}

Experiments are conducted in simulation environments by setting the maximum user interaction length to 30 and the exit mechanism parameter \( W \) to 3. The results are presented in Figure~\ref{fig:ablation}. \textbf{HRL4PFG-wo-hie} performs the worst in both item-side fairness and accuracy, demonstrating the necessity of hierarchical learning for fairness-aware recommendations. \textbf{HRL4PFG-wo-tc} outperforms \textbf{HRL4PFG-wo-hie} but remains inferior to the full model, indicating that imposing constraints on the fairness-guided target generated by the high-level agent can enhance model performance. \textbf{HRL4PFG-wo-fm} performs worse than \textbf{HRL4PFG}, suggesting that filtering items based on the fairness-guided target benefits the low-level agent's learning process. 

\subsection{Hyper-Parameter Sensitivity Analysis (RQ4)}
In this subsection, we examine the effects of \( \lambda_g \) in Equation \ref{eq:l_reward} and \( M \) in Equation \ref{eq:h_reward} on the model's performance. Here, \(\lambda_g\) controls the guidance strength applied to the low-level agent, and \(M\) determines the update interval for the high-level agent’s fairness-guided target.

\begin{figure}[h]
  {
    \centering
    \includegraphics[width=0.5 \textwidth]{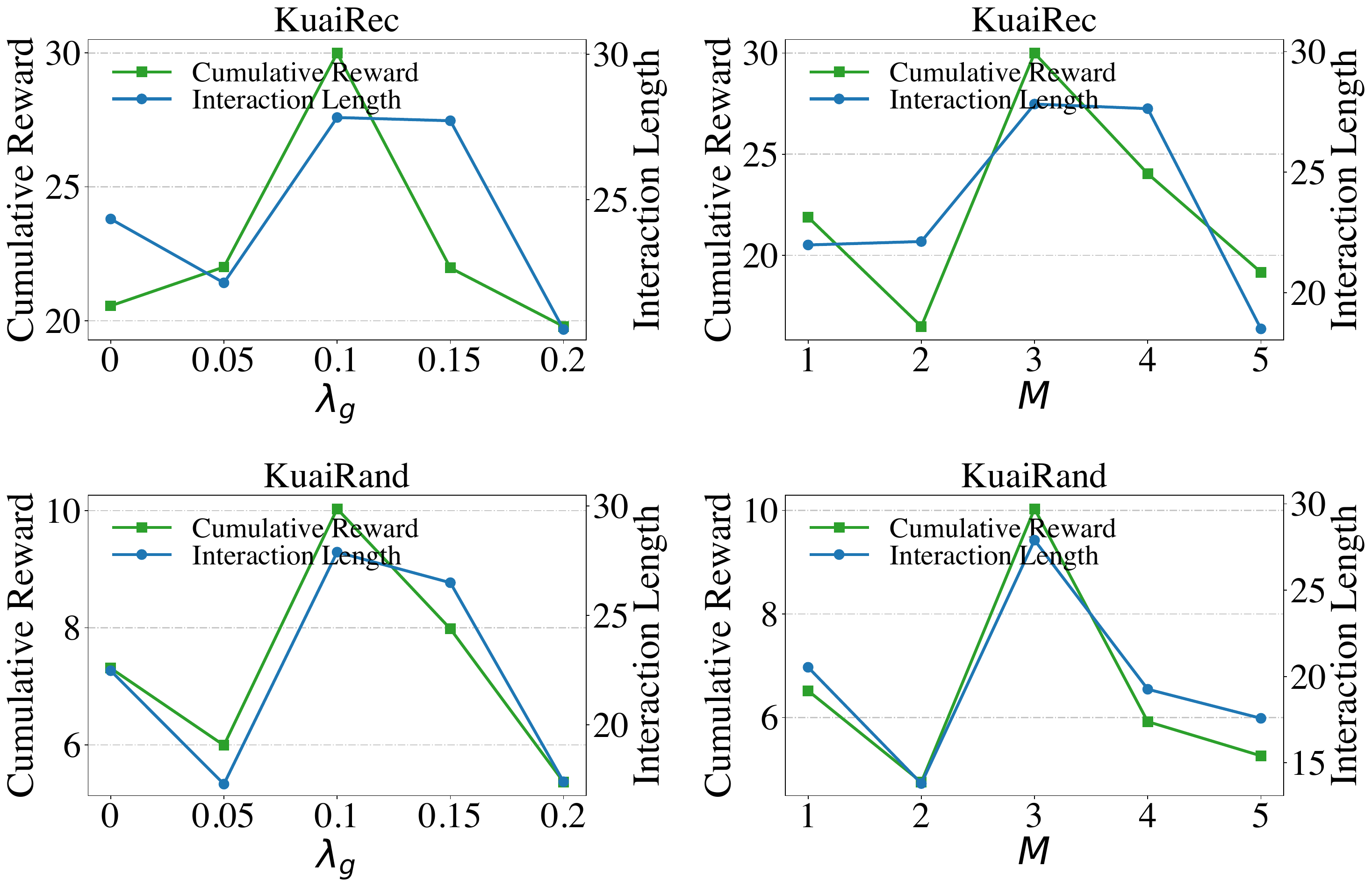}
    \caption{Hyper-parameter study on two datasets.}    
    \label{fig:hyper_parameter}
  }
\end{figure}

Experiments are conducted on the two dataset by setting the maximum interaction length to 30 and the exit mechanism parameter $W$ to 3. The results for various parameter settings are shown in Figure~\ref{fig:hyper_parameter}. As \(\lambda_g\) increases, both recommendation accuracy and fairness improve until \(\lambda_g = 0.1\), after which performance declines. This suggests that the fairness-guided target generated by the high-level agent should not excessively guide the decision-making process of the low-level agent within a single step. Similarly, increasing \(M\) improves performance up to \(M = 3\), beyond which the performance deteriorates, likely because too short an update interval lacks foresight and too long an interval makes the target too easily achieved, thereby diminishing the guiding effect.

\section{Conclusion}
In this paper, we presented HRL4PFG, a proactive guiding strategy for improving item-side fairness in interactive recommender systems. By using a hierarchical approach, HRL4PFG progressively guides user preference toward long-tail items, enhancing item-side fairness while maintaining user satisfaction. Experimental results show that HRL4PFG improves both item-side fairness and user satisfaction compared to existing methods.

Future directions include adapting HRL4PFG to other fairness goals like demographic fairness, integrating richer user behaviors for better adaptability, and conducting real-world studies to assess long-term effects on fairness and satisfaction

\bibliographystyle{named}
\bibliography{ijcai26}

\end{document}